\documentclass[aps,prd,twocolumn,showpacs,preprintnumbers,superscriptaddress]{revtex4-2}
\usepackage{orcidlink}
\usepackage[T1]{fontenc}
\usepackage[utf8]{inputenc}
\usepackage{amsmath,amssymb,amsfonts}
\usepackage{graphicx}
\usepackage{bm}
\usepackage{braket}
\usepackage{slashed}
\usepackage{xcolor}
\usepackage{hyperref}
\usepackage{subcaption}

\newcommand{\be}{\begin{equation}}
\newcommand{\ee}{\end{equation}}
\newcommand{\ba}{\begin{eqnarray}}
\newcommand{\ea}{\end{eqnarray}}

\begin{document}

\title{Diffractive Production of Heavy Quarkonia at the Electron Ion Collider}

\author{Mohammad Yousuf Jamal\orcidlink{0000-0003-2386-5755}}
\affiliation{Institute of Particle Physics and Key Laboratory of Quark and Lepton Physics (MOE),
Central China Normal University, Wuhan 430079, China}

\author{Ajaharul Islam\orcidlink{0000-0001-8174-907X}}
\affiliation{Institute of Particle Physics and Key Laboratory of Quark and Lepton Physics (MOE),
Central China Normal University, Wuhan 430079, China}

\author{Aritra Bandyopadhyay\orcidlink{0000-0002-3058-7258}}
\affiliation{Institute of Theoretical Physics, University of Wrocław, plac Maksa Borna 9, PL-50204 Wrocław, Poland}
\affiliation{Department of Physics, West University of Timişoara, Bd. Vasile Pârvan 4, Timişoara 300223, Romania}

\author{Santosh K. Das\orcidlink{0000-0003-3867-3158}}
\affiliation{School of Physical Sciences, Indian Institute of Technology Goa,
Ponda 403401, Goa, India}
%\date{\today}

\begin{abstract}
We study diffractive photo- and electroproduction of the $S$-wave heavy quarkonia
$J/\psi$, $\psi(2S)$, and $\Upsilon(nS)$ at energies relevant for the Electron-Ion
Collider (EIC). The production amplitude is evaluated while retaining the full
transverse-momentum ($\ell_t$) dependence of the hard two-gluon kernel, that is, without expanding
the impact-parameter Bessel kernel as is done in the small-size color-dipole limit. The quarkonia light-cone wave functions are built from Cornell-potential solutions of the
Schr\"odinger equation, normalized to the measured leptonic widths, and combined with a
modern collinear gluon distribution. After benchmarking the framework against the full
set of HERA charmonium cross-section ratio data, we provide a
consistent set of bottomonium cross-section ratio predictions in EIC kinematics. We find that the full
$\ell_t$-resolved treatment systematically improves the description of the radially
excited states relative to the leading dipole limit, and we identify the kinematic
windows where this difference is largest.
\end{abstract}


\maketitle

%% =====================================================================
\textit{Introduction.} Diffractive photo- and electroproduction of heavy vector mesons
has long been recognized as a clean probe of small-$x$ gluon dynamics. The large quark
mass provides a hard scale that justifies a perturbative treatment, while the exclusive
final state selects coherent two-gluon exchange from the target. In the leading
logarithmic approximation, the forward amplitude is proportional to the gluon momentum
density $xG(x,\mu^2)$, evaluated at a momentum fraction $x$ and a factorization scale
$\mu^2$ of order the heavy-quark mass squared. This proportionality yields the characteristic rise of the cross section with the photon-proton center-of-mass energy $W$, which was among the clearest signatures of gluon dominance at small Bjorken-$x$ observed at HERA~\cite{Ryskin:1992ui,Ryskin:1995hz,Frankfurt:1997fj,ZEUS:2002wfj}.

The forthcoming Electron-Ion Collider (EIC), operating at $\sqrt{s}\simeq 29$ to
$140$~GeV with high luminosity and excellent forward instrumentation, will turn these
measurements into precision imaging of gluons in protons and
nuclei~\cite{Abir:2023fpo, AbdulKhalek:2021gbh,Boer:2021ehu,Fanelli:2022kro}. It will reach $x$ as low
as $10^{-4}$ to $10^{-5}$ and span $Q^2$ from quasi-real photoproduction to several tens
of GeV$^2$, a regime where the gluon density remains only moderately constrained and
exclusive quarkonium production offers complementary sensitivity to
$xG(x,\mu^2)$~\cite{Martin:2007sb,Jones:2013pga}. Extending these studies to nuclear
targets will further probe gluon shadowing, saturation, and nuclear
geometry~\cite{Lappi:2013am,Tu:2020mvm}.

The standard theoretical tool is the color-dipole picture, in which the $q\bar{q}$ pair
produced by the photon is assumed compact relative to the scale of the target gluon
field~\cite{Nemchik:1997xb,Frankfurt:1997fj,Wusthoff:1999cr}. This description works well
for $J/\psi$ production~\cite{H1:2002yab} and has been applied extensively to coherent
and incoherent production off nuclei~\cite{Frankfurt:1998yf,Martin:2007sb}. For radially
excited states, however, it is more delicate, because the ratio
$\sigma(\psi(2S))/\sigma(J/\psi)$ and its analogues are governed by node-induced
cancellations in the $2S$ and $3S$ wave functions, which are sensitive to the full
distribution of dipole sizes and gluon
virtualities~\cite{Hoyer:1999xe,Cepila:2019skb,Peredo:2023oym}. The transverse separation
of the pair, $b$, and the gluon transverse momentum, $\ell_t$, enter the hard kernel
through the combination $\ell_t b$, and the dipole limit amounts to expanding the kernel
for $\ell_t b\ll 1$, which discards precisely the structure to which the excited states
are most sensitive. Dipole and saturation-based (BK) approaches give complementary predictions for these ratios~\cite{Peredo:2023oym, Mantysaari:2021ryb}. Here, we quantify the effects of the full gluon transverse-momentum dependence beyond the dipole approximation.

A treatment that retains the full $\ell_t$ dependence avoids this expansion. The
impact-parameter formalism of Refs.~\cite{Ryskin:1995hz,Frankfurt:1997fj,Suzuki:2000az}
provides a natural setting, in which the Bessel function $J_0(\ell_t b)$ is kept exact
and the dipole limit is recovered only upon its small-argument expansion. A realistic
quarkonium wave function with the correct radial-node structure is the second key
ingredient. Early analyses relied on simple light-cone ansatze or potential-model inputs
without a controlled nonrelativistic-to-light-cone
mapping~\cite{Buchmuller:1980su, Quigg:1977dd}, whereas more systematic constructions
start from a Cornell potential and perform the full transformation to light-cone
variables~\cite{Frankfurt:1997fj,Suzuki:2000az}. We build on that approach here.

In this letter, we combine Cornell-potential wave functions, normalized to the measured
leptonic widths $\Gamma_{ee}$, with a direct numerical evaluation of the full
$\ell_t$-dependent amplitude. For the gluon distribution we use the \textsc{\textsc{herapdf2.0} nlo} set
(variation series)~\cite{H1:2015ubc}, accessed through the \textsc{lhapdf}
library~\cite{Buckley:2014ana}, with an infrared matching scale
$Q_0^2 = 1.001~\mathrm{GeV}^2$. We benchmark the framework against HERA data for
$\sigma(\psi(2S))/\sigma(J/\psi)$ in $Q^2$, $W$, and $|t|$, and then present
bottomonium-ratio predictions in EIC kinematics. The chosen ranges $0 \le Q^2 \le 40~\mathrm{GeV}^2$, $40 \le W \le 200~\mathrm{GeV}$, and $0 \le |t| \le 1~\mathrm{GeV}^2$ are broadly compatible with the kinematic reach anticipated for the EIC. While the $Q^2$ and $|t|$ intervals lie well within the expected acceptance, the upper end of the $W$ range approaches that of the highest-energy EIC configurations. These ranges therefore cover the phase space most relevant for exclusive heavy-quarkonium production studies~\cite{Accardi:2012qut, AbdulKhalek:2021gbh}. Throughout, the full $\ell_t$-resolved
result is compared with the leading dipole limit obtained from the same framework and
the same inputs, so that the comparison isolates the effect of the $\ell_t$ expansion
alone. Predictions for $\Upsilon$ ratios in related approaches have appeared
before~\cite{GayDucati:2016ryh,Peredo:2023oym,Cepila:2019skb}, and our results provide an
independent, $\ell_t$-resolved determination obtained with potential-model wave functions
calibrated to the HERA charmonium data.

\begin{figure}[t]
\centering
\includegraphics[height=6cm,width=8cm]{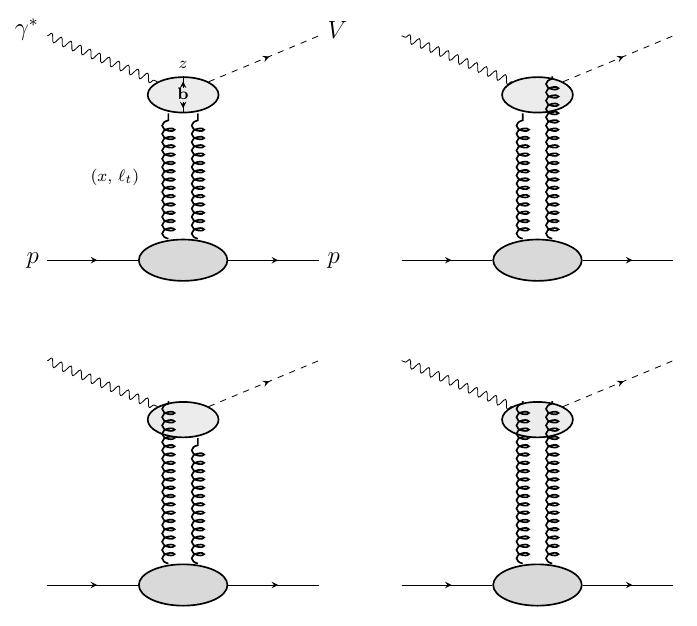}
\caption{Schematic of exclusive diffractive vector-meson production
$\gamma^*(Q^2)+p\to V+p$ via color-singlet two-gluon exchange. The four diagrams correspond to the possible attachments of the exchanged gluons to the quark–antiquark dipole generated by the virtual photon. Here $z$ denotes the quark light-cone momentum fraction, $b$ the transverse separation of the $q\bar q$ dipole, and $\ell_t$ the transverse momentum of the exchanged gluons.}
\label{fig:diagram}
\end{figure}

%% =====================================================================
\textit{Framework.} We consider the process $\gamma^*(Q^2)+p\to V+p$, shown
schematically in Fig.~\ref{fig:diagram}, at small Bjorken $x=(Q^2+4m_q^2)/W^2$ and for
$V=J/\psi,\psi(2S),\Upsilon(1S,2S,3S)$. The $z$-dependent forward amplitude is written as
\begin{equation}
\mathcal{A}(z;Q^2, W)
= \int d^2\mathbf{b}\;
\phi_\gamma(z,\mathbf{b};Q^2)\,
\mathcal{M}(x,\mathbf{b})\,
\psi_V(z,\mathbf{b}),
\label{eq:amp}
\end{equation}
where $z$ is the light-cone momentum fraction carried by the quark, $\mathbf{b}$ is the
transverse separation of the $q\bar{q}$ pair, $\phi_\gamma$ and $\psi_V$ are the photon
and vector meson light-cone (LC) wave functions, and $\mathcal{M}$ is the convolution kernel
that encodes the coupling to the gluon distribution through two-gluon exchange.

%% =====================================================================

\begin{figure*}[t]
\centering
\begin{subfigure}[t]{0.49\linewidth}
\includegraphics[height=5cm,width=8cm]{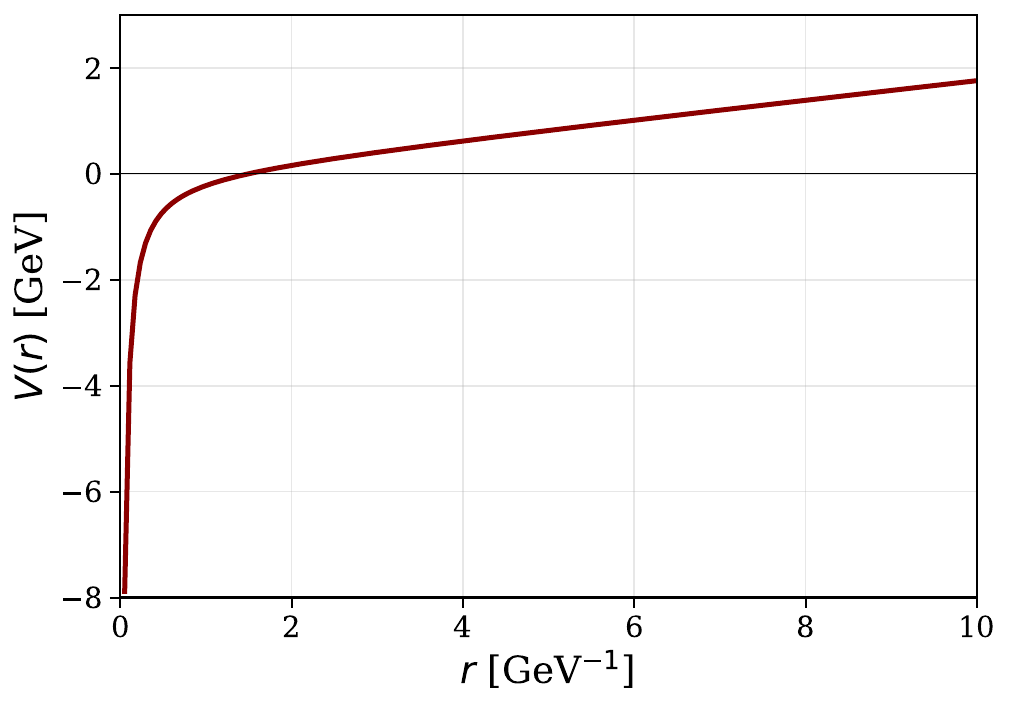}
\caption{}
\label{fig:cornell_potential}
\end{subfigure}\hfill
\begin{subfigure}[t]{0.49\linewidth}
\includegraphics[height=5cm,width=8cm]{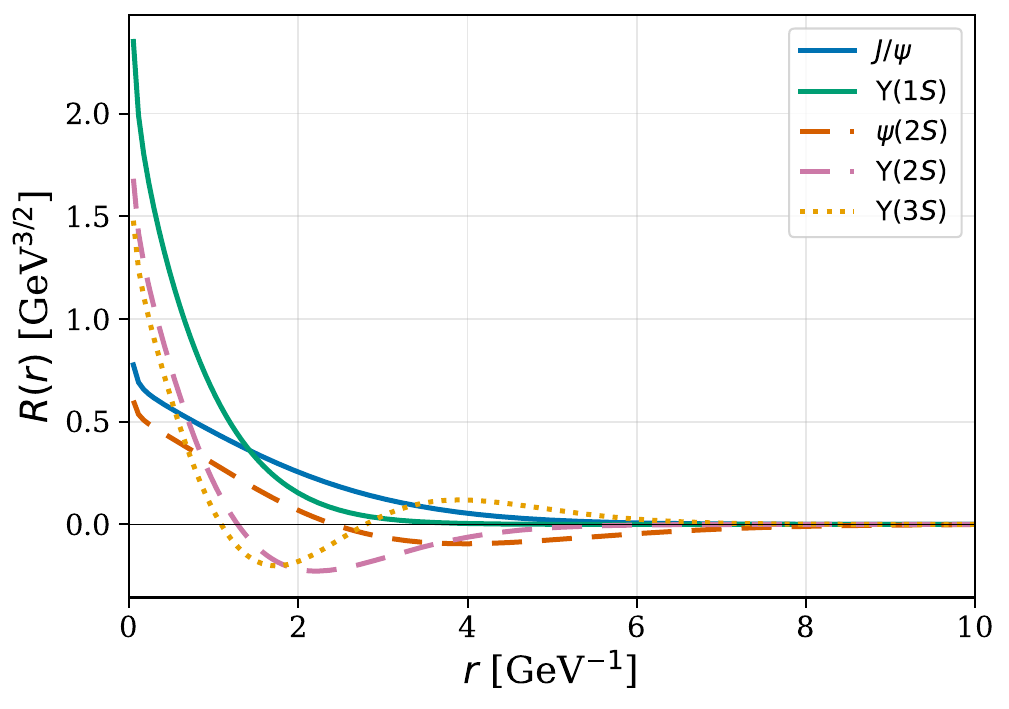}
\caption{}
\label{fig:radial_wavefunctions}
\end{subfigure}
\caption{ (a) Cornell potential of Eq.~\eqref{eq:corn} for $\alpha_s=0.30$ and
$\sigma=0.18~\mathrm{GeV}^2$.
(b) Radial wave functions $R_{nS}(r)$ for the charmonium states
$J/\psi$ and $\psi(2S)$, and the bottomonium states
$\Upsilon(1S)$, $\Upsilon(2S)$, and $\Upsilon(3S)$, obtained by solving the nonrelativistic Schr\"odinger equation with the Cornell potential.}
\label{fig:pot_wf}
\end{figure*}

\begin{figure*}[t]
\centering
\begin{subfigure}[t]{0.32\linewidth}
\includegraphics[height=5cm,width=6cm]{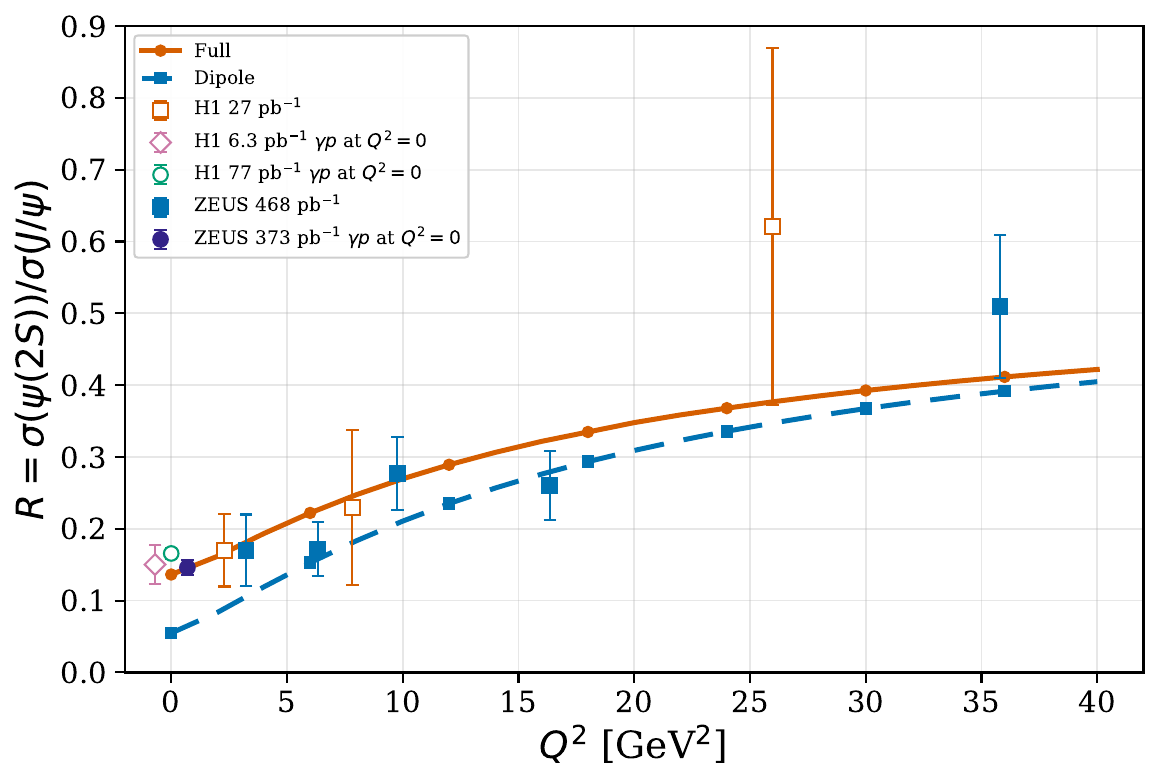}
\caption{}
\label{fig:R_vs_Q2}
\end{subfigure}\hfill
\begin{subfigure}[t]{0.32\linewidth}
\includegraphics[height=5cm,width=6cm]{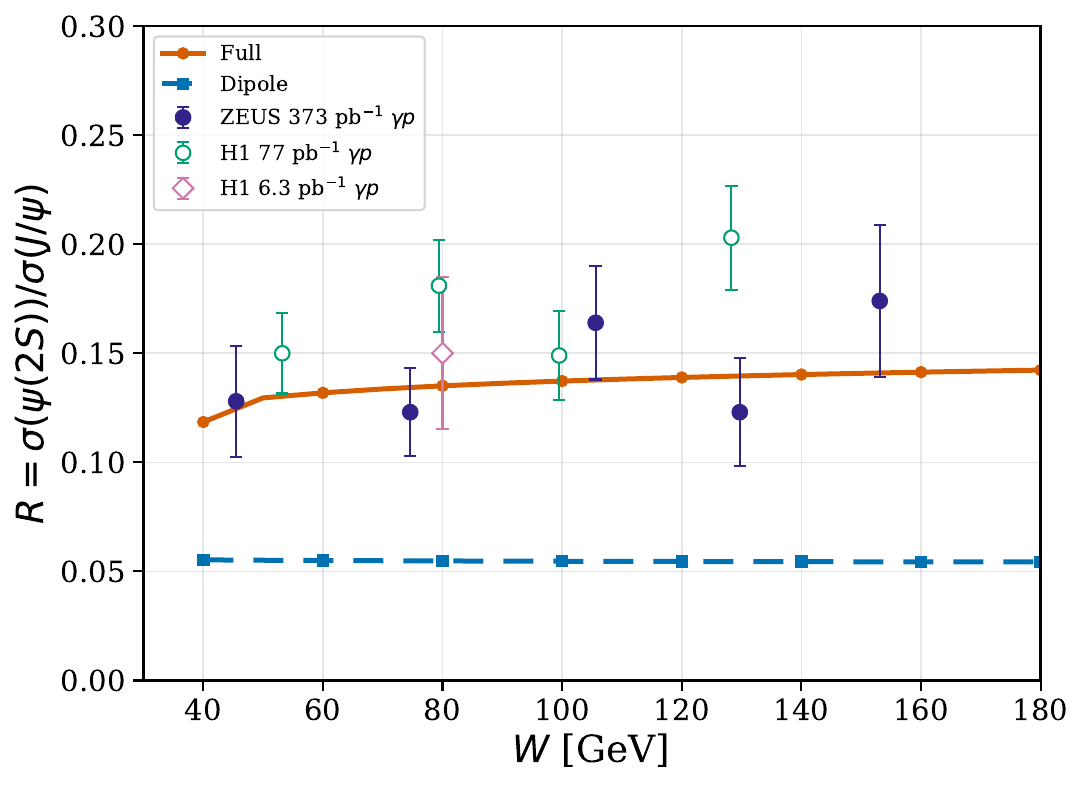}
\caption{}
\label{fig:R_vs_W}
\end{subfigure}\hfill
\begin{subfigure}[t]{0.32\linewidth}
\includegraphics[height=5cm,width=6cm]{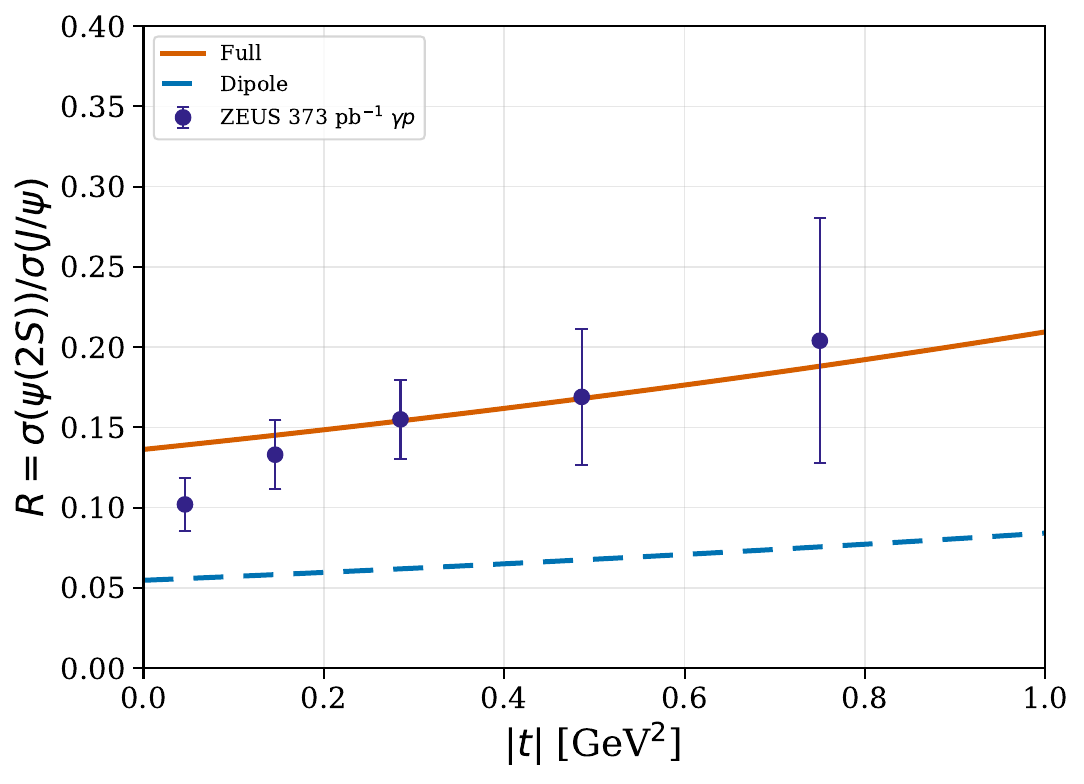}
\caption{}
\label{fig:R_vs_t}
\end{subfigure}
\caption{Ratio $R = \sigma(\psi(2S))/\sigma(J/\psi)$ in exclusive photo- and
electroproduction: (a) versus $Q^2$ at $W = 90$~GeV, (b) versus $W$ at $Q^2 = 0$, and
(c) versus $|t|$ at $W = 90$~GeV, $Q^2 = 0$. Solid (dashed) curves are the full (dipole)
results within the present framework. Data are from
ZEUS~\cite{ZEUS:2022mfb,ZEUS:2009asc} and H1~\cite{H1:2005dtp,H1:2002yab,H1:1997kxo}.}
\label{fig:R_psi2S}
\end{figure*}

\begin{figure*}[t]
\centering
\begin{subfigure}[t]{0.32\linewidth}
\includegraphics[height=5cm,width=6cm]{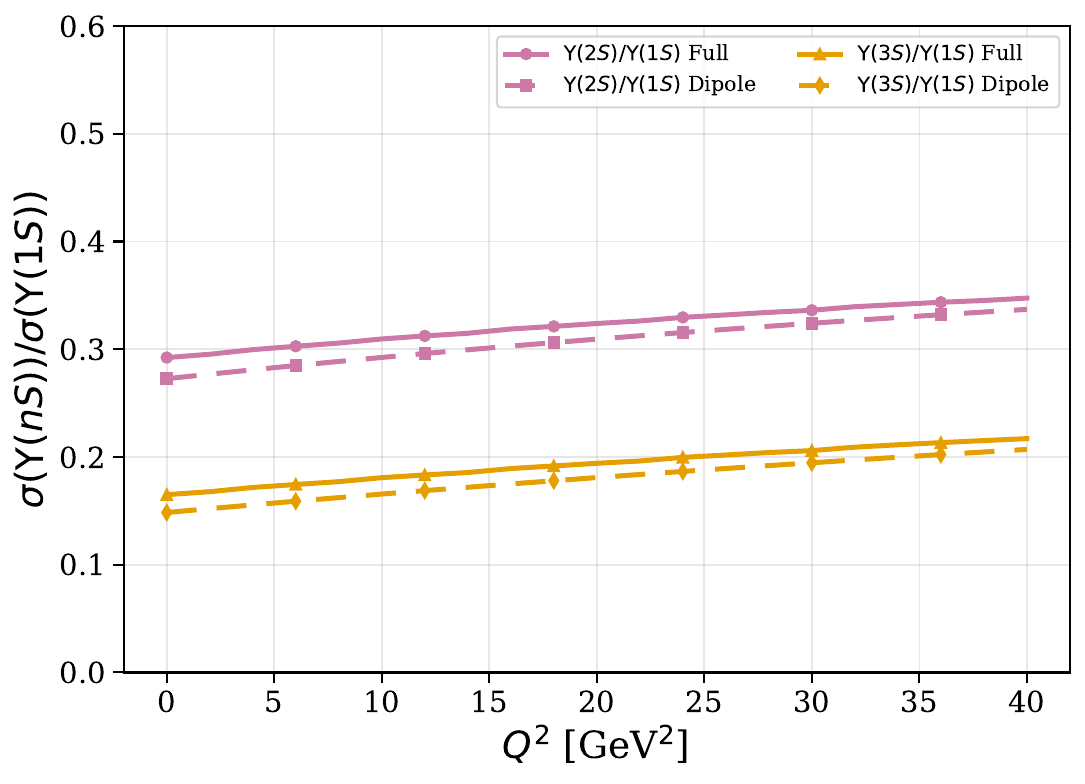}
\caption{}
\label{fig:Upsilon_vs_Q2}
\end{subfigure}\hfill
\begin{subfigure}[t]{0.32\linewidth}
\includegraphics[height=5cm,width=6cm]{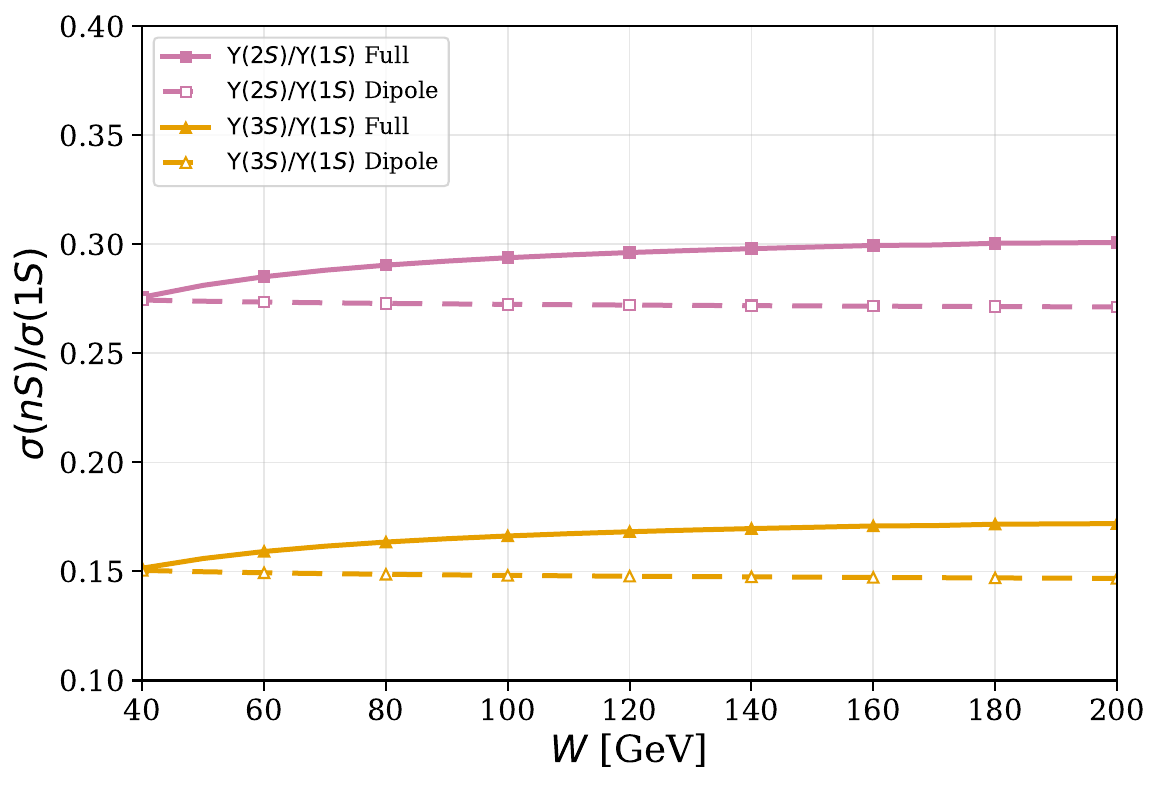}
\caption{}
\label{fig:Upsilon_vs_W}
\end{subfigure}\hfill
\begin{subfigure}[t]{0.32\linewidth}
\includegraphics[height=5cm,width=6cm]{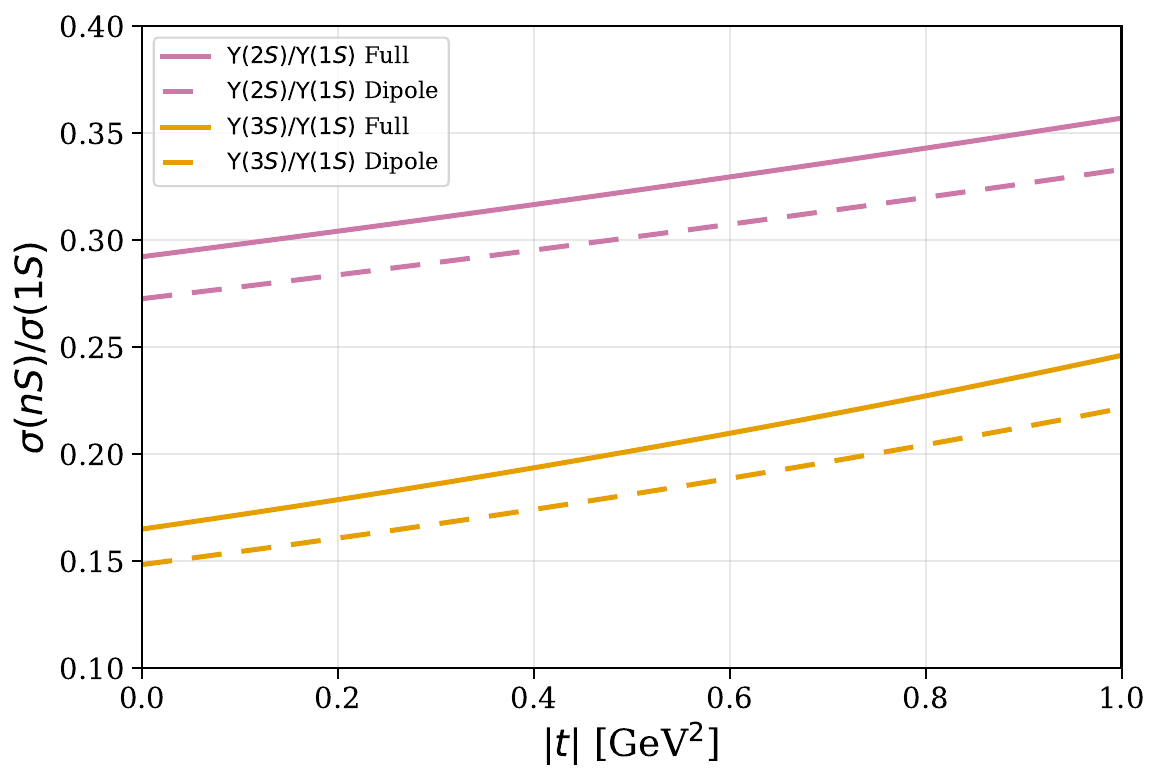}
\caption{}
\label{fig:Upsilon_vs_t}
\end{subfigure}
\caption{EIC predictions for $\sigma(\Upsilon(nS))/\sigma(\Upsilon(1S))$ ($n=2,3$):
(a) versus $Q^2$ at $W = 90$~GeV, (b) versus $W$ at $Q^2 = 0$, and (c) versus $|t|$ at
$W = 90$~GeV, $Q^2 = 0$. Solid (dashed) curves are the full (dipole) results.}
\label{fig:R_Upsilon}
\end{figure*}

\textit{Quarkonium wave functions:} We solve the $S$-wave radial Schr\"odinger equation
with a Cornell potential,
\begin{equation}
\mathcal{V}_C(r) = -\frac{C_F\alpha_s}{r} + \sigma r,
\label{eq:corn}
\end{equation}
with $C_F=4/3$, $\alpha_s=0.3$, and $\sigma=0.18$~GeV$^2$, obtaining radial wave
functions $R_{nS}(r)$ for charmonium and bottomonium. The short-distance normalization
is fixed by the measured leptonic width through
$\Gamma_{ee}(V\to e^+e^-) = 4\alpha_{\rm em}^2 e_Q^2 |R_V(0)|^2/M_V^2$, $M_V$ being the vector meson mass. The non-relativistic (NR) momentum-space wave functions are then mapped to LC momentum space using
\begin{equation}
\psi_V(z,k_t) = \left[\frac{k_t^2+m_q^2}{4[z(1-z)]^3}\right]^{1/4}
\psi_{\rm NR}\!\left(\sqrt{\frac{k_t^2+(2z-1)^2m_q^2}{4z(1-z)}}\right),
\label{eq:lcmap}
\end{equation}
and then transformed to impact-parameter space by a Hankel transform. This mapping is
the standard kinematical Terentev prescription and does not include the Melosh spin
rotation or the $D$-wave admixture generated by a massive photon-like $V\to Q\bar{Q}$ vertex, which are known to affect the $\psi(2S)/J/\psi$ ratio, in particular its $W$ dependence at low $Q^2$~\cite{Krelina:2019gee,Krelina:2019egg,Henkels:2020kju}. The sensitivity of excited to ground-state ratios to the wave-function model was also noted early on for $\Upsilon$ production~\cite{Rybarska:2008pk}. The Cornell input mitigates this model dependence, since it contains both the Coulomb short-distance and the linear confining behavior, its normalization is fixed by the measured leptonic widths, and realistic potentials give similar ratios~\cite{Cepila:2019skb}. Since the full and dipole results are computed with identical wave functions, the difference between them, which is the focus of this Letter, is insensitive to this choice. The photon LC wave function is $\phi_\gamma(z,b;Q^2)= K_0(b\,Q_{\rm eff})$, with $Q_{\rm eff}^2(z)=z(1-z)Q^2+m_q^2$. This transverse virtuality is the only hard scale in the process, so it also sets the upper limit of the gluon evolution below.

\textit{Full amplitude and dipole limit:} The central $\ell_t$-dependent overlap is
\begin{equation}
A(\ell_t,z;Q^2) = \frac{1}{\ell_t^2}\int_0^\infty db\,b\,
[1-J_0(\ell_t b)]\,\phi_\gamma(z,b;Q^2)\,\psi_V(z,b),
\label{eq:Alt}
\end{equation}
where $[1-J_0(\ell_t b)]/\ell_t^2 \to b^2/4$ as $\ell_t\to 0$. The full $z$-dependent
amplitude then reads
\begin{multline}
\mathcal{A}_{\rm full}(z;Q^2,W) = 4\pi\alpha_s\biggl\{xG(x,Q_0^2)\,A_0(z;Q^2) \\
+ \int_{Q_0^2}^{Q_{\rm eff}^2(z)} d\mu^2\,
\frac{\partial xG(x,\mu^2)}{\partial\mu^2}\,
A(\sqrt{\mu^2},z;Q^2)\biggr\},
\label{eq:full}
\end{multline}

where $A_0(z;Q^2) = \frac{1}{4}\int_0^\infty db\,b^3\,\phi_\gamma\,\psi_V$ is the
dipole-limit overlap and $Q_0^2$ is an infrared separation scale below which
$A(\ell_t,z;Q^2)\approx A_0(z;Q^2)$. The boundary term proportional to $xG(x,Q_0^2)$
resums the infrared region $\ell_t^2<Q_0^2$, while the integral builds up the
$\ell_t$-resolved contribution through the scale derivative $\partial xG/\partial\mu^2$.
The leading dipole approximation follows by replacing $A(\ell_t,z;Q^2)\to A_0(z;Q^2)$
throughout Eq.~(\ref{eq:full}), which collapses the amplitude to
\begin{equation}
\mathcal{A}_{\rm dipole}(z;Q^2,W) = 4\pi\alpha_s\,xG(x,Q_{\rm eff}^2)\,A_0(z;Q^2).
\label{eq:dipole}
\end{equation}
The convergence of the small-$\ell_t b$ expansion underlying Eq.~(\ref{eq:dipole}) is
examined quantitatively in the Supplemental Material~\cite{supp}, where we show that it
fails for the radially excited states.

\textit{From the amplitude to the cross section:} Finally, performing the $z$-integration requires classifying the longitudinal and transverse projections of the amplitude,
\begin{equation}
I_L = \int_0^1\! dz\;z(1-z)\,\mathcal{A}(z),\quad
I_T = \int_0^1\! dz\;\frac{\mathcal{A}(z)}{z(1-z)},
\end{equation}
which reflect the LC structure of the $\gamma^*\to q\bar{q}$ vertex, with longitudinal photons
weighting the amplitude near $z\simeq 1/2$ and transverse photons receiving enhanced
contributions from the endpoints $z\to 0,1$. The forward differential cross-section at
$t=0$ for polarization $\lambda=L,T$ is then
\begin{equation}
\left.\frac{d\sigma_{\lambda}}{dt}\right|_{t=0}
= \frac{\mathcal{N}^2}{16\pi}\,\bigl(1+\rho^2\bigr)\,I_\lambda^2(Q^2,W),
\label{eq:dsigdt}
\end{equation}
where $\rho$ is the ratio of the real to imaginary parts of the amplitude, estimated from
the standard logarithmic derivative, and $\mathcal{N}^2\propto\alpha_s^2 e_Q^2/M_V$
collects the overall normalization, including the skewness
correction~\cite{Martin:2007sb}. Assuming the standard exponential
$t$-dependence of diffractive vector-meson production at small $|t|$,
\begin{equation}
\frac{d\sigma}{dt} = \left.\frac{d\sigma}{dt}\right|_{t=0}\,e^{-B_V(W)\,|t|},
\label{eq:dsdt_exp}
\end{equation}
integration over $t$ gives 
\begin{equation}
    \sigma_\lambda(W,Q^2) = B_V^{-1}(W)\,\left.\frac{d\sigma_\lambda}{dt}\right|_{t=0},
\end{equation}
and the measured electroproduction cross
section is $\sigma = \sigma_T + \epsilon\,\sigma_L$, with $\epsilon$ the virtual-photon
polarization parameter. The diffractive slope follows the Regge parametrization
\begin{equation}
B_V(W) = B_0 + 4\,\alpha'\,\ln(W/W_0),
\label{eq:Bslope}
\end{equation}
with $W_0 = 90~\mathrm{GeV}$ and soft-Pomeron slope $\alpha' = 0.135~\mathrm{GeV}^{-2}$
from the H1 elastic $J/\psi$ analysis~\cite{H1:2013okq}. The intercepts $B_0$ are taken
from the corresponding HERA measurements, where available, and extrapolated by
transverse-size scaling for the higher bottomonium states, with the values used and a
test of their impact given in the Supplemental Material~\cite{supp}. In the cross-section
ratios studied below, the normalization $\mathcal{N}$, the skewness factor, and the slope
$B_V$ largely cancel, leaving the observables sensitive mainly to the interplay of the
wave function and the gluon density.

%% =====================================================================
\textit{Gluon input:} A central feature of this work is the use of a modern collinear
gluon PDF. The gluon momentum density $xG(x,\mu^2)$ is read directly from the \textsc{herapdf2.0}
NLO variation set~\cite{H1:2015ubc} through the \textsc{\textsc{lhapdf}6}
interface~\cite{Buckley:2014ana}, without any external re-parametrization. Because $xG$
enters both the boundary term and the evolution integral in Eq.~(\ref{eq:full}), the
scale derivative $\partial xG/\partial\mu^2$ controls how the different $\ell_t$ modes
contribute to the amplitude, and its shape directly affects the predicted $Q^2$
dependence of the ratios. The numerical procedure, the choice of $Q_0^2$, and the stability of the results against its variation are detailed in the Supplemental Material~\cite{supp}.

The scale structure of Eq.~(\ref{eq:full}) parallels the NLO collinear treatment of
Refs.~\cite{Jones:2013pga,Jones:2013eda}, where a Sudakov factor suppresses gluon
emissions with transverse momenta above a hard cutoff scale. At leading order their amplitude is proportional to the gluon density at a single effective scale, which corresponds to the dipole limit of Eq.~(\ref{eq:full}), while the Sudakov factor restores sensitivity to the gluon transverse-momentum distribution, in analogy with the explicit $\partial xG/\partial\mu^2$ integral retained here. The two treatments are therefore complementary probes of the same scale physics.

\textit{Results.} Figure~\ref{fig:pot_wf} shows the Cornell potential and the resulting
radial wave functions. The ground states are nodeless and spatially compact, with
bottomonium more localized than charmonium because of the heavier quark mass. The excited states display the expected radial nodes and larger transverse sizes; a quantitative convergence test of the dipole expansion, and three-dimensional visualizations of the key intermediate quantities are provided in the Supplemental Material~\cite{supp}.

Figure~\ref{fig:R_psi2S} shows the ratio $R=\sigma(\psi(2S))/\sigma(J/\psi)$ as a function
of $Q^2$, $W$, and $|t|$. The full calculation (solid curves) reproduces the mild $Q^2$
dependence and the overall normalization of the HERA data from H1 and
ZEUS~\cite{H1:2002yab,ZEUS:2022mfb,ZEUS:2009asc,H1:2005dtp,H1:1997kxo}, whereas the dipole
limit (dashed curves) lies systematically lower and exhibits a different $Q^2$ shape.
Since many systematic effects cancel in the ratio, this difference reflects the physics
of the $\ell_t$ expansion rather than a normalization choice. The same pattern holds for
the $W$ dependence at $Q^2=0$ in Fig.~\ref{fig:R_vs_W}, where the full result tracks the
data across the HERA energy range, and for the $|t|$ dependence in
Fig.~\ref{fig:R_vs_t}, where the dipole result shows a systematically harder slope.

The physical origin of this difference is the radial node of the $\psi(2S)$ wave
function, which sits near the typical transverse separation of the pair and produces a
partial cancellation between the inner and outer lobes of the overlap with the photon.
The size of that cancellation depends on the distribution of gluon virtualities probed,
which the dipole limit fixes to a single effective scale while the full calculation
retains in its entirety. This retained spread of scales is precisely what the ratio is
sensitive to, and it accounts for the cleaner description of the data in the full
treatment.

Figure~\ref{fig:R_Upsilon} presents the bottomonium ratios $\sigma(\Upsilon(nS))/\sigma(\Upsilon(1S))$ for $n=2,3$, for which we are not aware of any existing experimental data. The heavier $b$ mass reduces typical dipole sizes, so the node-induced cancellations are milder than for charmonium, yet the full and dipole results still differ appreciably over a broad $Q^2$ range, as seen in Fig.~\ref{fig:Upsilon_vs_Q2}. The full result is smoother and more monotonic, while the dipole limit shows a stronger $Q^2$ variation and a shifted
normalization. The $W$ dependence in Fig.~\ref{fig:Upsilon_vs_W} is nearly flat at
$Q^2=0$, with the $\Upsilon(3S)/\Upsilon(1S)$ ratio lying below
$\Upsilon(2S)/\Upsilon(1S)$ as a consequence of the stronger node suppression of the $3S$
state, and the $|t|$ dependence in Fig.~\ref{fig:Upsilon_vs_t} shows a mild rise in the
full calculation against the flatter dipole prediction. A precision EIC measurement of
these ratios would therefore provide a clean and direct test of the $\ell_t$-resolved
framework.

\textit{Conclusions.} We have computed diffractive heavy-quarkonium production amplitudes while retaining the full $\ell_t$ dependence of the hard kernel, using Cornell-potential LC wave functions normalized to leptonic widths, together with a modern collinear gluon distribution. Three conclusions emerge. First, the full $\ell_t$-resolved amplitude markedly improves the description of $R_{\psi(2S)}$ across $Q^2$, $W$, and $|t|$ relative to the leading dipole limit evaluated with identical inputs, where the node-induced cancellations in the $\psi(2S)$ wave function are essential. Second, even for bottomonium, where the dipole picture is expected to be more reliable, nontrivial differences between the full and dipole treatments persist and are accessible at the EIC. Third, the framework naturally interpolates between the dipole and $k_T$-factorization descriptions and can be extended to nuclear targets, offering a unified $\ell_t$-resolved description of exclusive quarkonium production in the EIC era. 

The $\gamma^{(*)} + p\to V + p$ amplitudes computed here are also the direct theoretical input for exclusive quarkonium production in ultraperipheral collisions at the LHC, where the $\psi(2S)/J/\psi$ ratio has recently been measured as a function of rapidity by LHCb~\cite{LHCb:2024pcz} and ALICE~\cite{ALICE:2026mlg}. An extension of the present framework to ultraperipheral kinematics is, therefore, a natural next step.

\begin{acknowledgments}
The authors are deeply grateful to Jan-e Alam and Bedangadas Mohanty for several fruitful discussions and valuable suggestions. M.Y.J and A.I. gratefully acknowledge Central China Normal University (CCNU), Wuhan, China, for its
hospitality and financial support through postdoctoral fellowships.
A.B. acknowledges support from the ULAM fellowship program of the Polish National Agency for Academic Exchange (NAWA), No.~BNI/ULM/2024/1/00193 and EU’s NextGenerationEU instrument through the National Recovery and Resilience Plan of Romania - Pillar III-C9-I8, managed by the Ministry of Research, Innovation and Digitization, within the project entitled ``Facets of Rotating Quark-Gluon Plasma'' (FORQ), contract no.~760079/23.05.2023 code CF 103/15.11.2022.
S.K.D. acknowledges the support from Anusandhan National Research Foundation (ANRF), India, under grant No.:ANRF/ARG/2025/002424/PS
\end{acknowledgments}

\bibliographystyle{apsrev4-2}
\bibliography{main}

@article{Suzuki:2000az,
    author = "Suzuki, K. and Hayashigaki, A. and Itakura, K. and Alam, J. and Hatsuda, T.",
    title = "{Validity of the color dipole approximation for diffractive production of heavy quarkonium}",
    eprint = "hep-ph/0005250",
    archivePrefix = "arXiv",
    doi = "10.1103/PhysRevD.62.031501",
    journal = "Phys. Rev. D",
    volume = "62",
    pages = "031501",
    year = "2000"
}

@article{H1:2013okq,
    author = "Alexa, C. and others",
    collaboration = "H1",
    title = "{Elastic and Proton-Dissociative Photoproduction of J/psi Mesons at HERA}",
    eprint = "1304.5162",
    archivePrefix = "arXiv",
    primaryClass = "hep-ex",
    reportNumber = "DESY-13-058",
    doi = "10.1140/epjc/s10052-013-2466-y",
    journal = "Eur. Phys. J. C",
    volume = "73",
    number = "6",
    pages = "2466",
    year = "2013"
}

@article{ZEUS:2022mfb,
    author = "Abt, I. and others",
    collaboration = "ZEUS",
    title = "{Measurement of exclusive photoproduction of $\psi(2S)$ mesons at HERA}",
    eprint = "2206.13343",
    archivePrefix = "arXiv",
    primaryClass = "hep-ex",
    reportNumber = "DESY-22-104",
    doi = "10.1007/JHEP12(2022)164",
    journal = "JHEP",
    volume = "12",
    pages = "164",
    year = "2022"
}

@article{ZEUS:2009asc,
    author = "Chekanov, S. and others",
    collaboration = "ZEUS",
    title = "{Exclusive electroproduction of $J/\psi$ mesons at HERA}",
    eprint = "hep-ex/0404008",
    archivePrefix = "arXiv",
    reportNumber = "DESY-04-039",
    doi = "10.1016/j.nuclphysb.2004.06.034",
    journal = "Nucl. Phys. B",
    volume = "695",
    pages = "3--37",
    year = "2004"
}

@article{H1:2005dtp,
    author = "Aktas, A. and others",
    collaboration = "H1",
    title = "{Elastic $J/\psi$ production at HERA}",
    eprint = "hep-ex/0510016",
    archivePrefix = "arXiv",
    reportNumber = "DESY-05-161",
    doi = "10.1140/epjc/s2005-02448-9",
    journal = "Eur. Phys. J. C",
    volume = "46",
    pages = "585--603",
    year = "2006"
}

@article{H1:1997kxo,
    author = "Adloff, C. and others",
    collaboration = "H1",
    title = "{Diffractive charmonium spectrum at high energies}",
    eprint = "hep-ex/9712020",
    archivePrefix = "arXiv",
    reportNumber = "DESY-97-228",
    doi = "10.1007/s100520050241",
    journal = "Eur. Phys. J. C",
    volume = "3",
    pages = "13--18",
    year = "1998"
}

@article{H1:2002yab,
    author = "Adloff, C. and others",
    collaboration = "H1",
    title = "{Diffractive photoproduction of psi(2S) mesons at HERA}",
    eprint = "hep-ex/0205107",
    archivePrefix = "arXiv",
    reportNumber = "DESY-02-075",
    doi = "10.1016/S0370-2693(02)02275-X",
    journal = "Phys. Lett. B",
    volume = "541",
    pages = "251--264",
    year = "2002"
}

@article{Ryskin:1992ui,
    author = "Ryskin, M. G.",
    title = "{Diffractive J / psi electroproduction in LLA QCD}",
    reportNumber = "LU-TP-92-12",
    doi = "10.1007/BF01555742",
    journal = "Z. Phys. C",
    volume = "57",
    pages = "89--92",
    year = "1993"
}

@article{Frankfurt:1997fj,
    author = "Frankfurt, Leonid and Koepf, Werner and Strikman, Mark",
    title = "{Diffractive heavy quarkonium photoproduction and electroproduction in QCD}",
    eprint = "hep-ph/9702216",
    archivePrefix = "arXiv",
    reportNumber = "OSU-97-0201, DESY-97-028",
    doi = "10.1103/PhysRevD.57.512",
    journal = "Phys. Rev. D",
    volume = "57",
    pages = "512--526",
    year = "1998"
}

@article{Ryskin:1995hz,
    author = "Ryskin, M. G. and Roberts, R. G. and Martin, Alan D. and Levin, E. M.",
    title = "{Diffractive J / psi photoproduction as a probe of the gluon density}",
    eprint = "hep-ph/9511228",
    archivePrefix = "arXiv",
    reportNumber = "DTP-95-96, CBPF-NF-079-95, RAL-TR-95-065",
    doi = "10.1007/s002880050547",
    journal = "Z. Phys. C",
    volume = "76",
    pages = "231--239",
    year = "1997"
}

@article{Rybarska:2008pk,
    author = "Rybarska, A. and Schafer, W. and Szczurek, A.",
    title = "{Exclusive photoproduction of Upsilon: From HERA to Tevatron}",
    eprint = "0805.0717",
    archivePrefix = "arXiv",
    primaryClass = "hep-ph",
    doi = "10.1016/j.physletb.2008.08.022",
    journal = "Phys. Lett. B",
    volume = "668",
    pages = "126--132",
    year = "2008"
}

@article{Henkels:2020kju,
    author = "Henkels, Cheryl and de Oliveira, Emmanuel G. and Pasechnik, Roman and Trebien, Haimon",
    title = "{Exclusive photoproduction of excited quarkonia in ultraperipheral collisions}",
    eprint = "2004.00607",
    archivePrefix = "arXiv",
    primaryClass = "hep-ph",
    doi = "10.1103/PhysRevD.102.014024",
    journal = "Phys. Rev. D",
    volume = "102",
    number = "1",
    pages = "014024",
    year = "2020"
}

@article{Krelina:2019gee,
    author = "Krelina, M. and Goncalves, V. P. and Cepila, J.",
    title = "{Coherent and incoherent vector meson electroproduction in the future electron-ion colliders: the hot-spot predictions}",
    eprint = "1905.06759",
    archivePrefix = "arXiv",
    primaryClass = "hep-ph",
    doi = "10.1016/j.nuclphysa.2019.06.009",
    journal = "Nucl. Phys. A",
    volume = "989",
    pages = "187--200",
    year = "2019"
}

@article{Jones:2013eda,
    author = "Jones, S. P. and Martin, A. D. and Ryskin, M. G. and Teubner, T.",
    title = "{Predictions of exclusive {\ensuremath{\psi}}(2S) production at the LHC}",
    eprint = "1312.6795",
    archivePrefix = "arXiv",
    primaryClass = "hep-ph",
    reportNumber = "IPPP-13-107, DCPT-13-214, LTH-996",
    doi = "10.1088/0954-3899/41/5/055009",
    journal = "J. Phys. G",
    volume = "41",
    pages = "055009",
    year = "2014"
}

@article{Jones:2013pga,
    author = "Jones, S. P. and Martin, A. D. and Ryskin, M. G. and Teubner, T.",
    title = "{Probes of the small $x$ gluon via exclusive $J/\psi$ and $\Upsilon$ production at HERA and the LHC}",
    eprint = "1307.7099",
    archivePrefix = "arXiv",
    primaryClass = "hep-ph",
    reportNumber = "IPPP-13-52, DCPT-13-104, LTH-980",
    doi = "10.1007/JHEP11(2013)085",
    journal = "JHEP",
    volume = "11",
    pages = "085",
    year = "2013"
}

@article{Fanelli:2022kro,
    author = "Fanelli, C. and others",
    title = "{AI-assisted optimization of the ECCE tracking system at the Electron Ion Collider}",
    eprint = "2205.09185",
    archivePrefix = "arXiv",
    primaryClass = "physics.ins-det",
    doi = "10.1016/j.nima.2022.167748",
    journal = "Nucl. Instrum. Meth. A",
    volume = "1047",
    pages = "167748",
    year = "2023"
}

@article{AbdulKhalek:2021gbh,
    author = "Abdul Khalek, R. and others",
    title = "{Science Requirements and Detector Concepts for the Electron-Ion Collider}: {EIC Yellow Report}",
    eprint = "2103.05419",
    archivePrefix = "arXiv",
    primaryClass = "physics.ins-det",
    reportNumber = "BNL-220990-2021-FORE, JLAB-PHY-21-3198, LA-UR-21-20953",
    doi = "10.1016/j.nuclphysa.2022.122447",
    journal = "Nucl. Phys. A",
    volume = "1026",
    pages = "122447",
    year = "2022"
}

@article{Boer:2021ehu,
    author = {Boer, Dani{\"e}l and Pisano, Cristian and Taels, Pieter},
    title = "{Extracting color octet NRQCD matrix elements from $J/\psi$ production at the EIC}",
    eprint = "2102.00003",
    archivePrefix = "arXiv",
    primaryClass = "hep-ph",
    doi = "10.1103/PhysRevD.103.074012",
    journal = "Phys. Rev. D",
    volume = "103",
    number = "7",
    pages = "074012",
    year = "2021"
}

@article{Lappi:2013am,
    author = "Lappi, T. and Mantysaari, H.",
    title = "{$J/|psi$ production in ultraperipheral Pb+Pb and $p$+Pb collisions at energies available at the CERN Large Hadron Collider}",
    eprint = "1301.4095",
    archivePrefix = "arXiv",
    primaryClass = "hep-ph",
    doi = "10.1103/PhysRevC.87.032201",
    journal = "Phys. Rev. C",
    volume = "87",
    number = "3",
    pages = "032201",
    year = "2013"
}

@article{ZEUS:2002wfj,
    author = "Chekanov, S. and others",
    collaboration = "ZEUS",
    title = "{Exclusive photoproduction of J / psi mesons at HERA}",
    eprint = "hep-ex/0201043",
    archivePrefix = "arXiv",
    reportNumber = "DESY-02-008",
    doi = "10.1007/s10052-002-0953-7",
    journal = "Eur. Phys. J. C",
    volume = "24",
    pages = "345--360",
    year = "2002"
}

@article{Tu:2020mvm,
    author = "Tu, Zhoudunming",
    collaboration = "STAR",
    title = "{Exclusive J/ photoproduction off deuteron in d+Au ultra-peripheral collisions at STAR}",
    eprint = "2009.04860",
    archivePrefix = "arXiv",
    primaryClass = "nucl-ex",
    doi = "10.22323/1.387.0100",
    journal = "PoS",
    volume = "HardProbes2020",
    pages = "100",
    year = "2021"
}

@article{Frankfurt:1998yf,
    author = "Frankfurt, L. L. and McDermott, M. F. and Strikman, M.",
    title = "{Diffractive photoproduction of $\upsilon$ at HERA}",
    eprint = "hep-ph/9812316",
    archivePrefix = "arXiv",
    reportNumber = "DESY-98-196, MC-TH-98-18",
    doi = "10.1088/1126-6708/1999/02/002",
    journal = "JHEP",
    volume = "02",
    pages = "002",
    year = "1999"
}

@article{Nemchik:1997xb,
    author = "Nemchik, J. and Nikolaev, Nikolai N. and Predazzi, E. and Zakharov, B. G. and Zoller, V. R.",
    title = "{The Diffraction cone for exclusive vector meson production in deep inelastic scattering}",
    eprint = "hep-ph/9712469",
    archivePrefix = "arXiv",
    reportNumber = "FZ-IKP-97-23",
    doi = "10.1134/1.558573",
    journal = "J. Exp. Theor. Phys.",
    volume = "86",
    pages = "1054--1073",
    year = "1998"
}

@article{Wusthoff:1999cr,
    author = "Wusthoff, M. and Martin, Alan D.",
    title = "{The QCD description of diffractive processes}",
    eprint = "hep-ph/9909362",
    archivePrefix = "arXiv",
    reportNumber = "DTP-99-78",
    doi = "10.1088/0954-3899/25/12/201",
    journal = "J. Phys. G",
    volume = "25",
    pages = "R309--R344",
    year = "1999"
}

@article{Martin:2007sb,
    author = "Martin, A. D. and Nockles, C. and Ryskin, Mikhail G. and Teubner, Thomas",
    title = "{Small x gluon from exclusive J/psi production}",
    eprint = "0709.4406",
    archivePrefix = "arXiv",
    primaryClass = "hep-ph",
    reportNumber = "IPPP-07-50, DCPT-07-100, LTH-755",
    doi = "10.1016/j.physletb.2008.02.067",
    journal = "Phys. Lett. B",
    volume = "662",
    pages = "252--258",
    year = "2008"
}

@article{Buchmuller:1980su,
    author = "Buchmuller, W. and Tye, S. H. H.",
    title = "{Quarkonia and Quantum Chromodynamics}",
    reportNumber = "FERMILAB-PUB-80-094-T",
    doi = "10.1103/PhysRevD.24.132",
    journal = "Phys. Rev. D",
    volume = "24",
    pages = "132",
    year = "1981"
}

@article{Quigg:1977dd,
    author = "Quigg, C. and Rosner, Jonathan L.",
    title = "{Quarkonium Level Spacings}",
    reportNumber = "FERMILAB-PUB-77-082-T",
    doi = "10.1016/0370-2693(77)90765-1",
    journal = "Phys. Lett. B",
    volume = "71",
    pages = "153--157",
    year = "1977"
}

@article{Peredo:2023oym,
    author = "Peredo, Marco Alcazar and Hentschinski, Martin",
    title = "{Ratio of J/{\ensuremath{\Psi}} and {\ensuremath{\Psi}}(2s) exclusive photoproduction cross sections as an indicator for the presence of nonlinear QCD evolution}",
    eprint = "2308.15430",
    archivePrefix = "arXiv",
    primaryClass = "hep-ph",
    doi = "10.1103/PhysRevD.109.014032",
    journal = "Phys. Rev. D",
    volume = "109",
    number = "1",
    pages = "014032",
    year = "2024"
}

@article{Mantysaari:2021ryb,
    author = {M{\"a}ntysaari, Heikki and Penttala, Jani},
    title = "{Exclusive heavy vector meson production at next-to-leading order in the dipole picture}",
    eprint = "2104.02349",
    archivePrefix = "arXiv",
    primaryClass = "hep-ph",
    doi = "10.1016/j.physletb.2021.136723",
    journal = "Phys. Lett. B",
    volume = "823",
    pages = "136723",
    year = "2021"
}

@article{Hou:2019efy,
    author = "Hou, Tie-Jiun and others",
    title = "{New CTEQ global analysis of quantum chromodynamics with high-precision data from the LHC}",
    eprint = "1912.10053",
    archivePrefix = "arXiv",
    primaryClass = "hep-ph",
    reportNumber = "MSUHEP-19-025, PITT-PACC-1911, SMU-HEP-19-03",
    doi = "10.1103/PhysRevD.103.014013",
    journal = "Phys. Rev. D",
    volume = "103",
    number = "1",
    pages = "014013",
    year = "2021"
}

@article{Accardi:2012qut,
    author = "Accardi, A. and others",
    editor = "Deshpande, A. and Meziani, Z. E. and Qiu, J. W.",
    title = "{Electron Ion Collider: The Next QCD Frontier}: {Understanding the glue that binds us all}",
    eprint = "1212.1701",
    archivePrefix = "arXiv",
    primaryClass = "nucl-ex",
    reportNumber = "BNL-98815-2012-JA, JLAB-PHY-12-1652",
    doi = "10.1140/epja/i2016-16268-9",
    journal = "Eur. Phys. J. A",
    volume = "52",
    number = "9",
    pages = "268",
    year = "2016"
}

@misc{supp,
    note = "See Supplemental Material at [URL] for parameters, numerical details, the convergence analysis of the small-$\ell_t b$ expansion, and three-dimensional visualizations of the intermediate quantities, which includes Refs.\cite{Suzuki:2000az,Frankfurt:1997fj,ParticleDataGroup:2024cfk,H1:2013okq,H1:2013okq,H1:2002yab,H1:2015ubc,Buckley:2014ana,Hou:2019efy}."
}

@article{ParticleDataGroup:2024cfk,
    author = "Navas, S. and others",
    collaboration = "Particle Data Group",
    title = "{Review of particle physics}",
    doi = "10.1103/PhysRevD.110.030001",
    journal = "Phys. Rev. D",
    volume = "110",
    number = "3",
    pages = "030001",
    year = "2024"
}

@article{Abir:2023fpo,
    author = "Abir, Raktim and others",
    title = "{The case for an EIC Theory Alliance: Theoretical Challenges of the EIC}",
    journal = "DOI:",
    eprint = "2305.14572",
    archivePrefix = "arXiv",
    primaryClass = "hep-ph",
    reportNumber = "FERMILAB-FN-1226-T-V",
    doi = "10.2172/1975510",
    month = "5",
    year = "2023"
}

@article{ALICE:2026mlg,
    author = "Abdallah, Dana Ali Hassan and others",
    collaboration = "ALICE",
    title = "{Exclusive dimuon production and coherent charmonium photoproduction at forward rapidity in ultra-peripheral Pb$-$Pb collisions at $\sqrt{s_{\rm NN}}=5.36$ TeV}",
    journal = "arXiv e-prints",
    eprint = "2605.13569",
    archivePrefix = "arXiv",
    primaryClass = "nucl-ex",
    reportNumber = "CERN-EP-2026-143",
    year = "2026"
}

@article{LHCb:2024pcz,
    author = "Aaij, Roel and others",
    collaboration = "LHCb",
    title = "{Measurement of exclusive $J/\psi$ and $\psi(2S)$ production at $\sqrt{s}=13$ TeV}",
    eprint = "2409.03496",
    archivePrefix = "arXiv",
    primaryClass = "hep-ex",
    reportNumber = "LHCb-PAPER-2024-012, CERN-EP-2024-213",
    doi = "10.21468/SciPostPhys.18.2.071",
    journal = "SciPost Phys.",
    volume = "18",
    number = "2",
    pages = "071",
    year = "2025"
}

@article{GayDucati:2016ryh,
    author = "Gay Ducati, M. B. and Kopp, F. and Machado, M. V. T. and Martins, S.",
    title = "{Photoproduction of Upsilon states in ultraperipheral collisions at the CERN Large Hadron Collider within the color dipole approach}",
    eprint = "1610.06647",
    archivePrefix = "arXiv",
    primaryClass = "hep-ph",
    doi = "10.1103/PhysRevD.94.094023",
    journal = "Phys. Rev. D",
    volume = "94",
    number = "9",
    pages = "094023",
    year = "2016"
}

@article{Cepila:2019skb,
    author = "Cepila, Jan and Nemchik, Jan and Krelina, Michal and Pasechnik, Roman",
    title = "{Theoretical uncertainties in exclusive electroproduction of S-wave heavy quarkonia}",
    eprint = "1901.02664",
    archivePrefix = "arXiv",
    primaryClass = "hep-ph",
    doi = "10.1140/epjc/s10052-019-7016-9",
    journal = "Eur. Phys. J. C",
    volume = "79",
    number = "6",
    pages = "495",
    year = "2019"
}

@article{Hoyer:1999xe,
    author = "Hoyer, Paul and Peigne, Stephane",
    title = "{$J^\prime / \psi^\prime$ to $J/\psi$ ratio in diffractive photoproduction}",
    eprint = "hep-ph/9909519",
    archivePrefix = "arXiv",
    reportNumber = "NORDITA-1999-59-HE, LAPTH-749-99",
    doi = "10.1103/PhysRevD.61.031501",
    journal = "Phys. Rev. D",
    volume = "61",
    pages = "031501",
    year = "2000"
}

@article{H1:2015ubc,
    author = "Abramowicz, H. and others",
    collaboration = "H1, ZEUS",
    title = "{Combination of measurements of inclusive deep inelastic ${e^{\pm }p}$ scattering cross sections and QCD analysis of HERA data}",
    eprint = "1506.06042",
    archivePrefix = "arXiv",
    primaryClass = "hep-ex",
    reportNumber = "DESY-15-039",
    doi = "10.1140/epjc/s10052-015-3710-4",
    journal = "Eur. Phys. J. C",
    volume = "75",
    number = "12",
    pages = "580",
    year = "2015"
}

@article{Buckley:2014ana,
  author = {Buckley, Andy and others},
  title = {LHAPDF6: parton density access in the LHC precision era},
  journal = {Eur. Phys. J. C},
  volume = {75},
  year = {2015},
  number = {3},
  pages = {132},
  doi = {10.1140/epjc/s10052-015-3318-8}
}

@article{Krelina:2019egg,
    author = "Krelina, Michal and Nemchik, Jan and Pasechnik, Roman",
    title = "{$D$-wave effects in diffractive electroproduction of heavy quarkonia from the photon-like $V\rightarrow Q{\bar{Q}}$ transition}",
    eprint = "1909.12770",
    archivePrefix = "arXiv",
    primaryClass = "hep-ph",
    doi = "10.1140/epjc/s10052-020-7678-3",
    journal = "Eur. Phys. J. C",
    volume = "80",
    number = "2",
    pages = "92",
    year = "2020"
}

%\newpage
\begin{center}
{\Large\bf Supplemental Material}
\end{center}
This Supplemental Material collects the inputs, the numerical procedures, and the
auxiliary results referenced in the main text. Section~\ref{sec:params} lists the
parameters and the diffractive-slope intercepts, Section~\ref{sec:numerics} describes the numerical evaluation of the amplitude and the gluon-density input, Section~\ref{sec:dxg} explains the scale derivative of the gluon density, Section~\ref{sec:expansion} quantifies the convergence of the small-$\ell_t b$ expansion that underlies the dipole limit, and Section~\ref{sec:visual} presents three-dimensional
visualizations of the intermediate quantities.

%% =====================================================================
\section{Parameters and diffractive slopes}
\label{sec:params}

The Cornell potential uses $C_F=4/3$, $\alpha_s=0.30$, and $\sigma=0.18~\mathrm{GeV}^2$,
with heavy-quark masses $m_c=1.35$~GeV and $m_b=4.70$~GeV, consistent with the
potential-model construction of Refs.~\cite{Suzuki:2000az,Frankfurt:1997fj}. The leptonic
widths $\Gamma_{ee}$ that fix the normalization $|R_V(0)|^2$ are taken from the Particle
Data Group~\cite{ParticleDataGroup:2024cfk} and are listed in Table~\ref{tab:inputs}.

\begin{table}[h]
\centering
\caption{Vector-meson masses $M_V$ and leptonic widths $\Gamma_{ee}$ used to fix the
short-distance normalization $|R_V(0)|^2$, together with the diffractive-slope intercepts
$B_0$ at $W_0=90$~GeV used in Eq.~(11) of the main text.}
\label{tab:inputs}
\begin{tabular}{lccc}
\hline\hline
State & $M_V$ [GeV] & $\Gamma_{ee}$ [keV] & $B_0$ [GeV$^{-2}$] \\
\hline
$J/\psi$        & 3.0969  & 5.55  & 4.73 \\
$\psi(2S)$      & 3.6861  & 2.33  & 4.30 \\
$\Upsilon(1S)$  & 9.4603  & 1.340 & 4.00 \\
$\Upsilon(2S)$  & 10.0233 & 0.612 & 3.80 \\
$\Upsilon(3S)$  & 10.3552 & 0.443 & 3.60 \\
\hline\hline
\end{tabular}
\end{table}

The soft-Pomeron slope $\alpha'=0.135~\mathrm{GeV}^{-2}$, taken from the H1 elastic
$J/\psi$ analysis~\cite{H1:2013okq}, is the most precise determination available and is
used for all states, since no direct measurement exists for the excited charmonium and
bottomonium states and the soft-Pomeron slope is expected to be flavor-universal. The
$J/\psi$ and $\psi(2S)$ intercepts are anchored by H1
measurements~\cite{H1:2013okq,H1:2002yab}, while the $\Upsilon$ intercepts, which have no
direct measurement, are taken to decrease mildly with increasing meson mass following the
expected transverse-size scaling. Because the slope cancels almost completely in the
cross-section ratios, the precise $B_0$ values have little effect on the results reported
in the main text. Varying the $\Upsilon$ intercepts within $\pm 0.5~\mathrm{GeV}^{-2}$
changes the predicted ratios by less than a few percent.

%% =====================================================================
\section{Numerical evaluation and gluon input}
\label{sec:numerics}

The radial Schr\"odinger equation is solved on a logarithmic grid using the renormalized
Numerov method. The nonrelativistic wave function is mapped to light-cone momentum space
through Eq.~(3) of the main text and transformed to impact-parameter space by a Hankel
transform on a grid extending to $b_{\rm max}=12~\mathrm{GeV}^{-1}$, which is sufficient
for the $b^3$, $b^5$, and $b^7$ moments entering the overlap to converge.

The gluon momentum density $xG(x,\mu^2)$ is read directly from the \textsc{herapdf2.0 nlo}
variation set~\cite{H1:2015ubc} through the \textsc{lhapdf6} interface~\cite{Buckley:2014ana}. The
scale derivative $\partial xG/\partial\mu^2$ in Eq.~(5) of the main text is computed by a
symmetric finite difference at relative step $\varepsilon_{\rm rel}=5\times10^{-3}$, with
all scales clipped to the native PDF support. The infrared matching scale is set to
$Q_0^2=1.001~\mathrm{GeV}^2$, just above the lower edge of the \textsc{herapdf2.0} support
$\mu^2_{\rm min}=1.0~\mathrm{GeV}^2$. We have verified that the cross-section ratios are
stable against reasonable variation of $Q_0^2$ in the perturbative region, since the
boundary term proportional to $xG(x,Q_0^2)$ and the evolution integral combine to
reconstruct the full amplitude. We note that, at the lowest scales and the smallest $x$,
the \textsc{herapdf2.0} gluon density becomes very small, which makes the absolute forward cross
section for charmonium unreliable at the highest $W$. This feature largely cancels in the
cross-section ratios studied in the main text, since the same factor enters the numerator
and the denominator.

We have also quantified the sensitivity of the cross-section ratios to the gluon input. Varying the gluon over the full \textsc{herapdf2.0 nlo} variation set~\cite{H1:2015ubc}
changes the $\psi(2S)/J/\psi$ ratio by less than one percent across the entire $Q^2$
range, and repeating the calculation with an independent PDF family
(\textsc{ct18lo})~\cite{Hou:2019efy} leaves the ratio nearly unchanged. The gluon density
is least constrained at small $x$ and low scales, and this uncertainty affects the
absolute cross sections, but it largely cancels in the ratio, which is therefore the more
robust observable.

\begin{figure}[h]
\centering
\begin{subfigure}[t]{0.49\linewidth}
\includegraphics[height=3.5cm,width=4cm]{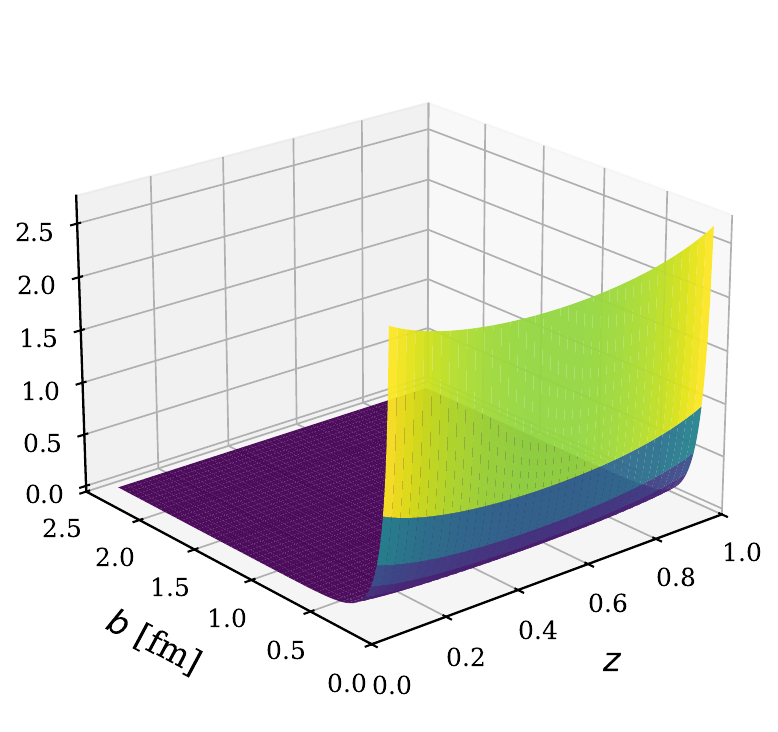}
\caption{}
\end{subfigure}\hfill
\begin{subfigure}[t]{0.49\linewidth}
\includegraphics[height=3.5cm,width=4cm]{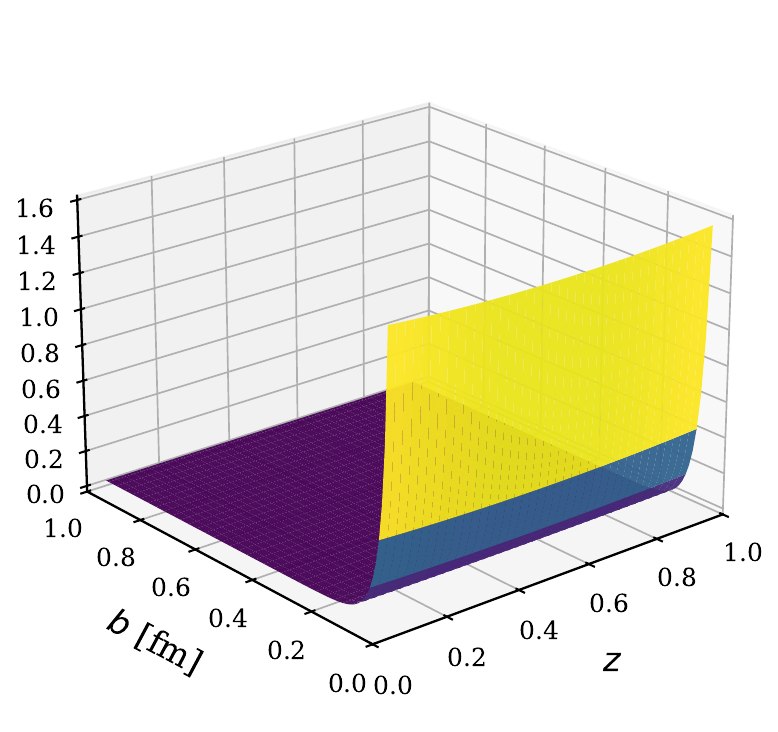}
\end{subfigure}

\begin{subfigure}[t]{0.49\linewidth}
\includegraphics[height=3.5cm,width=4cm]{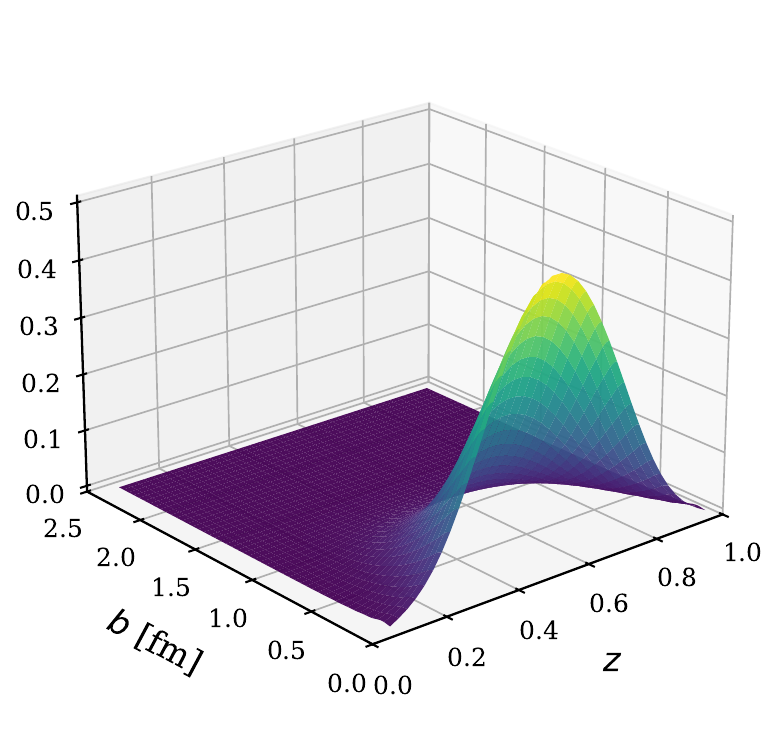}
\caption{}
\end{subfigure}\hfill
\begin{subfigure}[t]{0.49\linewidth}
\includegraphics[height=3.5cm,width=4cm]{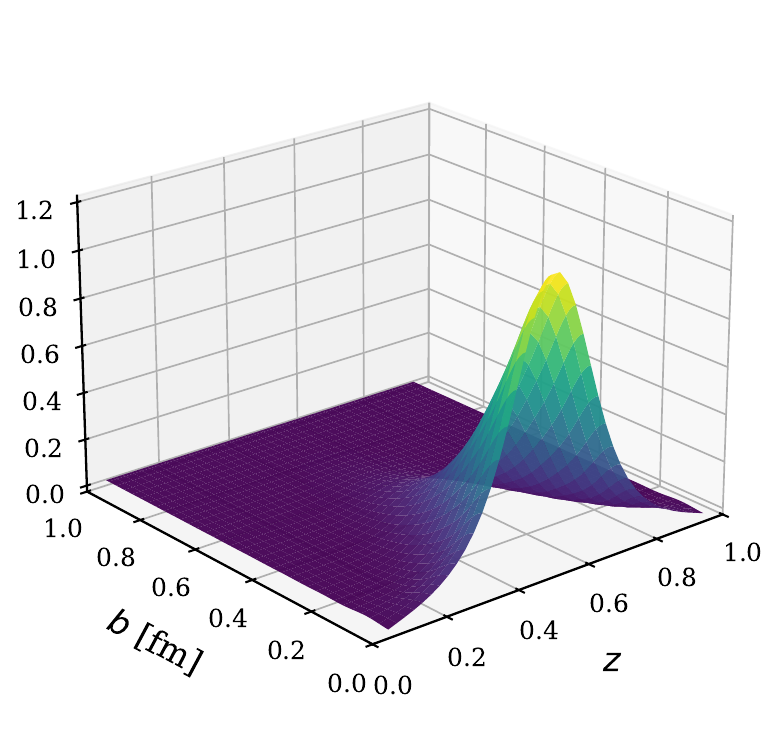}
\end{subfigure}
\caption{(a) Photon LC wave function $\phi_\gamma(z,b;Q^2)$ for the charm sector (left)
and the bottom sector (right). (b) Meson LC wave function $\psi_V(z,b)$ for $J/\psi$
(left) and $\Upsilon(1S)$ (right).}
\label{fig:phi_psi}
\end{figure}

\begin{figure}[h]
\centering
\begin{subfigure}[t]{0.49\linewidth}
\includegraphics[height=3.5cm,width=4cm]{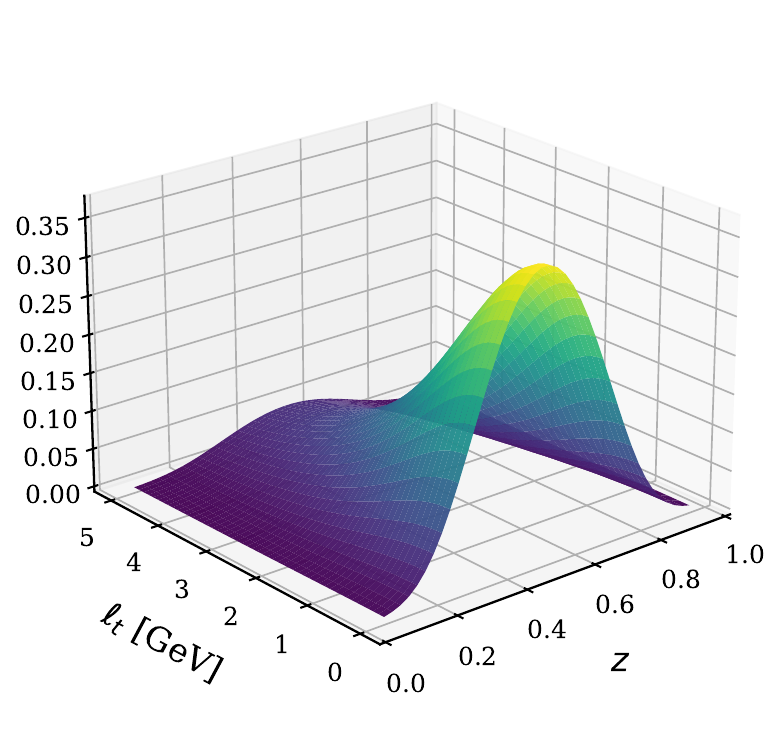}
\caption{}
\end{subfigure}\hfill
\begin{subfigure}[t]{0.49\linewidth}
\includegraphics[height=3.5cm,width=4cm]{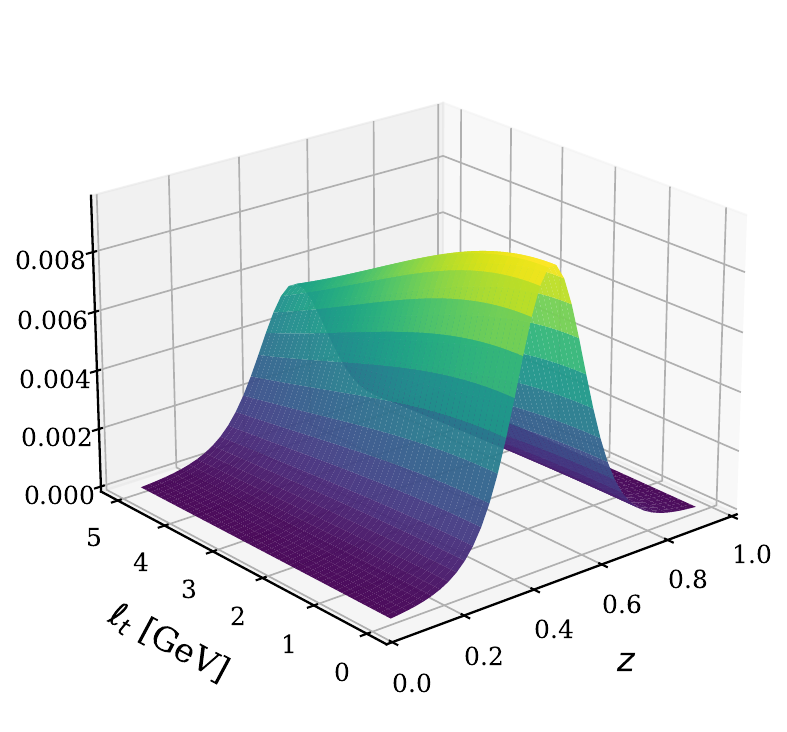}
\caption{}
\end{subfigure}
\caption{The $\ell_t$-dependent overlap $A(\ell_t,z;Q^2)$ at $Q^2=0$ for (a) $J/\psi$
and (b) $\Upsilon(1S)$.}
\label{fig:Alt}
\end{figure}

\begin{figure}[h]
\centering
\begin{subfigure}[t]{0.49\linewidth}
\includegraphics[height=3.5cm,width=4cm]{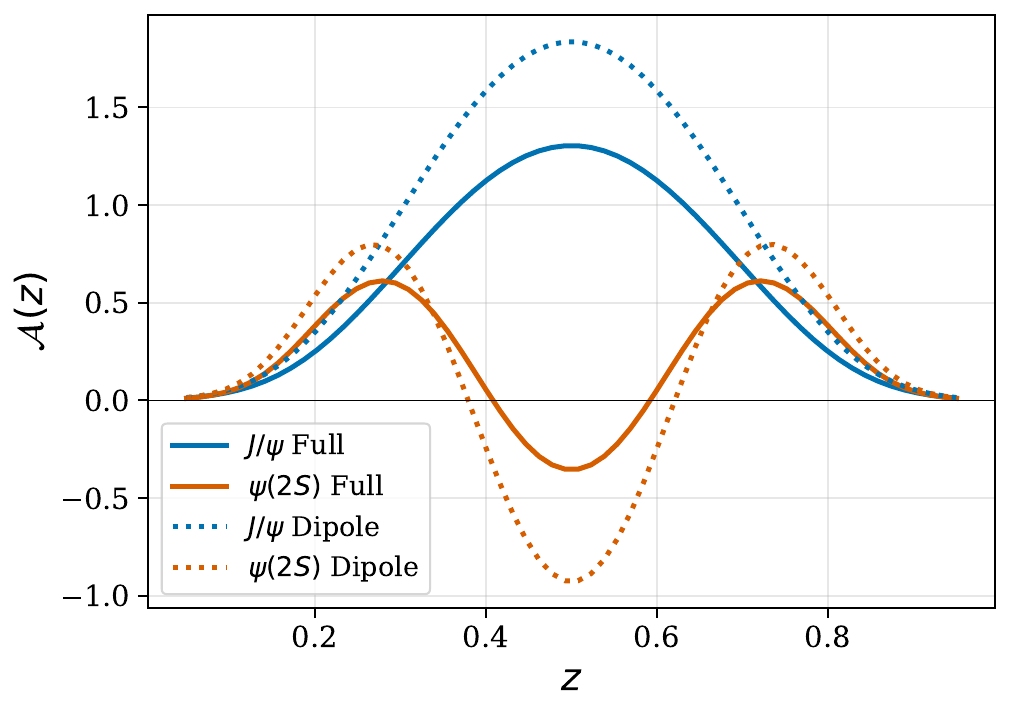}
\caption{}
\end{subfigure}\hfill
\begin{subfigure}[t]{0.49\linewidth}
\includegraphics[height=3.5cm,width=4cm]{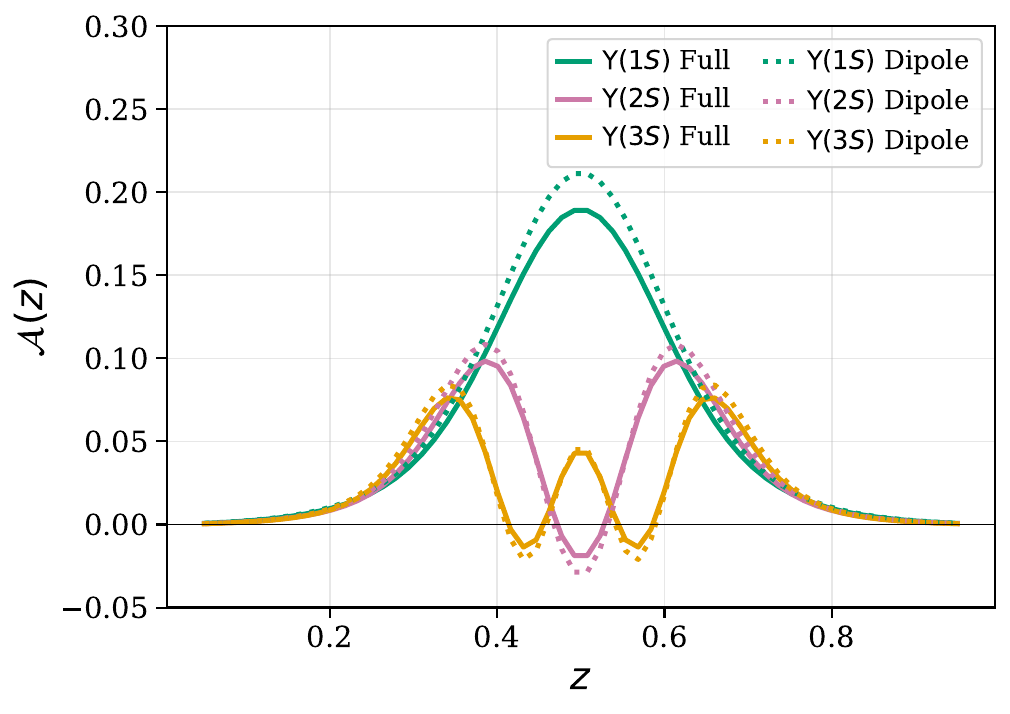}
\caption{}
\end{subfigure}
\caption{The $z$-dependent amplitude $\mathcal{A}(z;Q^2=0,W=90~\mathrm{GeV})$ in the full
(solid) and dipole (dotted) treatments for (a) charmonium and (b) bottomonium. The
node-induced sign change of the $\psi(2S)$ amplitude is visible.}
\label{fig:Az}
\end{figure}

%% =====================================================================
\section{The scale derivative of the gluon density}
\label{sec:dxg}

The full amplitude in Eq.~(5) of the main text contains the scale derivative
$\partial xG(x,\mu^2)/\partial\mu^2$ inside the evolution integral. This quantity has a
direct physical meaning. It measures the rate at which gluons are resolved as the
factorization scale increases. In this section we explain its role and describe how it
is evaluated numerically.

The starting point is the two-gluon exchange amplitude in the $k_t$-factorization form.
The coupling to the target proceeds through the unintegrated gluon distribution
$f(x,\ell_t^2)$, which is related to the collinear gluon density by
\begin{equation}
f(x,\ell_t^2) = \frac{\partial\, xG(x,\ell_t^2)}{\partial \ln \ell_t^2}
             = \ell_t^2\,\frac{\partial\, xG(x,\ell_t^2)}{\partial \ell_t^2}.
\label{eq:unint}
\end{equation}
The unintegrated distribution therefore carries the same information as the slope of the
collinear gluon in the scale variable. Writing the amplitude in terms of
$\partial xG/\partial\mu^2$ makes the connection to standard collinear PDF sets explicit,
since these sets provide $xG(x,\mu^2)$ on an interpolation grid.

Inserting Eq.~(\ref{eq:unint}) into the impact-parameter overlap and separating the
infrared region gives the structure used in Eq.~(5) of the main text. The contribution
below the matching scale $Q_0^2$ is resummed into the boundary term proportional to
$xG(x,Q_0^2)$. The contribution above $Q_0^2$ is built up by integrating the scale
derivative against the $\ell_t$-resolved overlap $A(\sqrt{\mu^2},z;Q^2)$. The upper limit
of this integral is the transverse virtuality $Q_{\rm eff}^2(z)=z(1-z)Q^2+m_q^2$, which is
the only hard scale in the process.

The role of the scale derivative is now transparent. In the dipole limit the overlap
$A(\ell_t,z;Q^2)$ is replaced by its $\ell_t\to 0$ value $A_0(z;Q^2)$, which is
independent of $\mu^2$. The evolution integral then collapses,
\begin{multline}
\int_{Q_0^2}^{Q_{\rm eff}^2} d\mu^2\,
\frac{\partial xG(x,\mu^2)}{\partial\mu^2}\,A_0(z;Q^2)= \\
\bigl[xG(x,Q_{\rm eff}^2)-xG(x,Q_0^2)\bigr]\,A_0(z;Q^2),
\end{multline}
and combining with the boundary term reproduces the dipole result
$\mathcal{A}_{\rm dipole}=4\pi\alpha_s\,xG(x,Q_{\rm eff}^2)\,A_0(z;Q^2)$ of Eq.~(6) of the
main text. The full calculation differs because $A(\sqrt{\mu^2},z;Q^2)$ varies with
$\mu^2$ across the integration range. The shape of $\partial xG/\partial\mu^2$ therefore
weights the different $\ell_t$ modes, and it controls how strongly the full result
departs from the dipole limit.

Numerically, the scale derivative is evaluated by a symmetric finite difference,
\begin{equation}
\frac{\partial xG(x,\mu^2)}{\partial\mu^2} \approx
\frac{xG(x,\mu^2(1+\varepsilon_{\rm rel})) - xG(x,\mu^2(1-\varepsilon_{\rm rel}))}
     {2\,\varepsilon_{\rm rel}\,\mu^2},
\label{eq:findiff}
\end{equation}
with relative step $\varepsilon_{\rm rel}=5\times10^{-3}$. The gluon density
$xG(x,\mu^2)$ is read directly from the \textsc{\textsc{\textsc{herapdf2.0} nlo}} interpolation
grid~\cite{H1:2015ubc} through the \textsc{\textsc{lhapdf6}} interface~\cite{Buckley:2014ana}. All scales
are clipped to the native support of the set. The \textsc{\textsc{herapdf2.0}} grid is sufficiently smooth in $\mu^2$ that the finite-difference result is insensitive to the precise choice of $\varepsilon_{\rm rel}$ over more than an order of magnitude around the adopted value $\varepsilon_{\rm rel} = 5\times10^{-3}$.

%% =====================================================================
\section{Convergence of the small-$\ell_t b$ expansion}
\label{sec:expansion}

The dipole limit in Eq.~(6) of the main text follows from expanding the
impact-parameter kernel for small $\ell_t b$,
\begin{equation}
\frac{1-J_0(\ell_t b)}{\ell_t^2}
= \frac{b^2}{4} - \frac{\ell_t^2 b^4}{64} + \frac{\ell_t^4 b^6}{2304} - \cdots,
\label{eq:bessel_expansion}
\end{equation}
and keeping only the leading $b^2$ term. The successive corrections involve the higher
$b$ moments
\begin{equation}
I_n(z) = \int_0^\infty db\, b^{\,n}\, \phi_\gamma(z,b)\,\psi_V(z,b),
\end{equation}
weighted by the corresponding moments of $\partial xG/\partial\mu^2$, and the expansion
is controlled only when $\ell_t b\ll 1$ over the region where the integrand has support.

For ground states the wave function $\psi_V$ is compact and the expansion converges
quickly. For radially excited states the outer lobe of $\psi_V$ extends to
$b\sim 3$ to $4~\mathrm{GeV}^{-1}$, while the relevant gluon transverse momenta are
$\ell_t\sim 1~\mathrm{GeV}$, so that $\ell_t b\sim 3$ and the expansion does not converge.
Table~\ref{tab:convergence} illustrates this for the $\psi(2S)$ and $J/\psi$ overlaps at
$z=0.5$, $Q^2=0$, where the higher $b$ moments of the $\psi(2S)$ exceed the leading term
in magnitude, confirming that the dipole expansion is uncontrolled for excited
charmonium. The full $\ell_t$-resolved amplitude, which keeps $1-J_0(\ell_t b)$ exact,
remains well defined because no expansion is involved.

\begin{table}[h]
\centering
\caption{Leading dipole overlap $A_0$ and the higher $b$ moments $I_5$ and $I_7$ at
$z=0.5$, $Q^2=0$, which control the $O(\ell_t^4)$ and $O(\ell_t^6)$ corrections, in
arbitrary common units. For the $\psi(2S)$ the higher moments exceed the leading term in
magnitude, signaling the breakdown of the small-$\ell_t b$ expansion.}
\label{tab:convergence}
\begin{tabular}{lccc}
\hline\hline
State & $A_0$ & $I_5$ & $I_7$ \\
\hline
$J/\psi$   & $+0.073$ & $+7.29$  & $+58.2$ \\
$\psi(2S)$ & $-0.041$ & $-11.22$ & $-166.7$ \\
\hline\hline
\end{tabular}
\end{table}

%% =====================================================================
\section{Three-dimensional visualizations}
\label{sec:visual}

To aid interpretation, we provide three-dimensional surfaces of the key intermediate
quantities. Figure~\ref{fig:phi_psi} shows the photon LC wave function
$\phi_\gamma(z,b;Q^2)$ and the meson LC wave function $\psi_V(z,b)$ for representative
charmonium and bottomonium states, Fig.~\ref{fig:Alt} shows the $\ell_t$-dependent overlap
$A(\ell_t,z;Q^2)$ for $J/\psi$ and $\Upsilon(1S)$, and Fig.~\ref{fig:Az} shows the full
$z$-dependent amplitude $\mathcal{A}(z;Q^2,W)$ compared between the full and dipole
treatments. These surfaces make explicit the node-induced sign changes in the
excited-state amplitudes and the kinematic regions where the full and dipole calculations
diverge most strongly.

%\bibliographystyle{apsrev4-2}
%\bibliography{main}

\end{document}